Obituary

In memoriam Stefan U. Egelhaaf
(17.06.1963 – 22.11.2023)

Authored by

M. A. Escobedo-Sánchez, M. Laurati, H. Löwen, W. C. K. Poon, P. N. Pusey, P. Schurtenberger

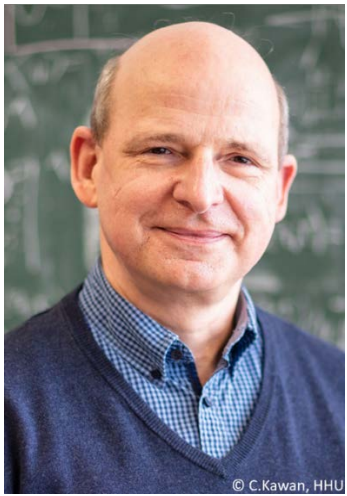

Stefan U. Egelhaaf (1963 – 2023)

We mourn the loss of an outstanding scientist and wonderful person, who passed away much too soon in November last year: Stefan Ulrich Egelhaaf. Stefan was born on 17th June 1963 in Schaffhausen (Switzerland) close to the German border. From 1988 to 1991 he studied physics and pharmacy at the Universities of Tübingen and Kiel. In Kiel, he finished his physics diploma with Klaus Schätzel in 1991 with a topic on light scattering. In 1995 Stefan received his PhD in biology from ETH Zürich. From 1995 to 1998 he was an instrument scientist at the Institute Laue Langevin in Grenoble, working on the D22 beamline, where he pioneered time-resolved small-angle neutron scattering experiments for the investigation of shape and phase transitions in surfactant systems. Moving to the School of Physics and Astronomy at the University of Edinburgh in 1998, he stayed until 2004, progressing from Lecturer to Reader to Professor. Stefan was appointed to a chair in "experimental physics of condensed matter" at the Heinrich-Heine University Düsseldorf in 2004 where he stayed ever since.

Stefan Egelhaaf's scientific contributions spanned the whole field of soft matter. His main areas of research concerned the non-equilibrium behaviour of colloidal dispersions, micelles and biological macromolecules. Focussing on the glass transition, gel formation and crystallization, he particularly investigated these phenomena under external influences and stimulating fields. He used a broad spectrum of experimental methods, ranging from static and dynamic light [1,2] and neutron [3,4] scattering to confocal [5] and digital Fourier microscopy [6,7]. Stefan's versatility and breadth of interests were demonstrated by highlights such as the discovery of a new glass state in strongly attractive particles (the so-called "attractive glass") [8]; the experimental proof of equilibrium cluster formation in colloidal fluids [9]; phase behavior of protein solutions [10]; the controlled realization of random optical energy landscapes and their influence on the dynamics of colloids [11]; and a fundamental understanding of sheared systems of soft condensed matter [12,13,14].

Throughout his research career, Stefan strived for deep understanding of complex problems, going for carefully designed ad executed experiments often combined with insightful theoretical work, rather than focussing on the quick publication of headline-grabbing results in "high-impact journals". Even as he progressed through his academic career, Stefan remaind a hands-on experimentalist in the lab and was eager to develop innovative experimental techniques. In preparing manuscripts Stefan spent much effort in reviewing and optimizing the presentation and language, because as he said: "If we are not accurate in presenting our findings, how can people trust that we are accurate in performing experiments?". It is therefore unfortunate for the soft matter community that Stefan was not able to fully realize his original ideas of "smart" adaptive colloids that perform tasks in a controlled manner using optical feedback and other external stimuli to unify colloid physics with artificial intelligence.

Stefan was an exceptional lecturer and an enthusiastic and inspiring teacher, going out of his way to find highly pedagogical means of putting across difficult subjects, and always showing his never-ending joy for science. His summer school lectures were legendary. During the last months of his life, he was updating several chapters for a new edition of the book based on the Bombannes "Scattering by Soft Matter" summer schools. Stefan had a keen eye for details that he used in his passion for teaching. He would add a ruler in PowerPoint to make every sketch, figure, symbol, and math font in his presentations, seminars or informal talks. Moreover, he would take the time to pick the proper colour coding to enhance attention to particular points in his discussion. He even used colour chalk while handwriting on the chalkboard in the lecture hall.

Stefan's conference presentations were always a joy to listen to. Whenever he spoke about a complex subject, it would be clear to the audience that he knew every detail of what he was presenting, and that the results reported were wholly trustworthy. At the same time, the way in which Stefan presented the findings was so clear, intuitive and passionate that even non-experts could easily follow the arguments and get excited about the topic.

In his research group, Stefan always kept his door open and welcomed everyone with a warm smile, an attentive ear and a sharp mind. With everyone he spoke to, Stefan would always try his best to meet them at their own level, because he was very conscious that everyone's path can be represented as a ladder where one needs to take one step at a time.

Last, but perhaps most importantly, Stefan was a wonderful and warm-hearted person. He had an infectious smile and humour without any trace of professorial arrogance; he was a true friend with whom one could not only have a scientific exchange, but could also enjoy laughs and jokes. Stefan was at the same time always gentle and calm, enthusiastic, pleasant and funny, a real friend who was always willing to help with advice and guidance.

Those of us who have had the privilege of knowing and working with Stefan personally will miss him terribly, and will remember him as a brilliant, enthusiastic, sincere, humorous, and loyal colleague and friend. We are all immensely grateful for the time we had with him. With his passing, the soft matter community has lost a dedicated researcher and teacher, and a highly-esteemed colleague.

Manuel A. Escobedo-Sánchez, Institut für Experimentelle Physik der kondensierten Materie, Universitätsstraße 1, Heinrich-Heine-Universität, 40225 Düsseldorf, Germany. E-Mail: escobedo@hhu.de

Marco Laurati, Department of Chemistry "Ugo Schiff", University of Florence, 50019 Sesto Fiorentino, Italy. E-Mail: marco.laurati@unifi.it

Hartmut Löwen, Institut für Theoretische Physik II: Weiche Materie, Heinrich-Heine-Universität, Universitätsstraße 1, 40225 Düsseldorf, Germany. E-Mail: hlowen@hhu.de

Wilson C.K. Poon, School of Physics & Astronomy, The University of Edinburgh, Edinburgh, EH9 3FD, UK. E-Mail: W.Poon@ed.ac.uk

Peter N. Pusey, School of Physics & Astronomy, The University of Edinburgh, Edinburgh, EH9 3FD, UK. E-Mail: Pusey@ed.ac.uk

Peter Schurtenberger, Division for Physical Chemistry, Lund University, Naturvetarvägen 14, 22100 Lund, Sweden. E-Mail: peter.schurtenberger@fkem1.lu.se